\documentclass[twocolumn,secnumarabic,amssymb, nobibnotes, aps, prd]{revtex4-1}
\usepackage{amsmath} \usepackage{amssymb}
\usepackage{graphicx} 
\usepackage{epstopdf}
\usepackage{amsmath}
\usepackage{amsmath,amssymb,amsthm,amsfonts,mathrsfs,bm,verbatim}
\usepackage{graphicx,subfigure}

\newcommand{\bea}{\begin{eqnarray}}
\newcommand{\eea}{\end{eqnarray}}
\def\nn{\nonumber}

\begin{document}

\title{Extremal rotating black holes, scalar perturbation and superradiant stability }

\author{Jia-Mao Lin}
\affiliation{Guangdong Provincial Key Laboratory of Quantum Engineering and Quantum Materials,\\ School of Physics and Telecommunication Engineering,\\
South China Normal University,Guangzhou 510006,China}

\author{Ming-Jian Luo}
\affiliation{Guangdong Provincial Key Laboratory of Quantum Engineering and Quantum Materials,\\ School of Physics and Telecommunication Engineering,\\
South China Normal University,Guangzhou 510006,China}
\author{Zi-Han Zheng}
\affiliation{Guangdong Provincial Key Laboratory of Quantum Engineering and Quantum Materials,\\ School of Physics and Telecommunication Engineering,\\
South China Normal University,Guangzhou 510006,China}
\author{Lei Yin}
\affiliation{Guangdong Provincial Key Laboratory of Nuclear Science, Institute of quantum matter, South China Normal University, Guangzhou 510006, China}
\affiliation{Guangdong-Hong Kong Joint Laboratory of Quantum Matter, Southern Nuclear Science Computing Center, South China Normal University, Guangzhou 510006, China}

\author{Jia-Hui Huang}
\email{huangjh@m.scnu.edu.cn}
\affiliation{Guangdong Provincial Key Laboratory of Quantum Engineering and Quantum Materials,
School of Physics and Telecommunication Engineering,
South China Normal University, Guangzhou 510006, China}
\affiliation{Institute of quantum matter, South China Normal University, Guangzhou 510006, China}
\affiliation{Guangdong-Hong Kong Joint Laboratory of Quantum Matter, Southern Nuclear Science Computing Center, South China Normal University, Guangzhou 510006, China}

\begin{abstract}
A (charged) rotating black hole may be unstable against a (charged) massive scalar field perturbation due to the existence of superradiance modes. The stability property depends on the parameters of the system. In this paper, the superradiant stable parameter space is studied for the four-dimensional extremal Kerr and Kerr-Newman black holes under massive and charged massive scalar perturbation.  For the extremal Kerr case, it is found that when the angular frequency and proper mass of the scalar perturbation satisfy the inequality $\omega<\mu/\sqrt{3}$, the extremal Kerr black hole and scalar perturbation system is superradiantly stable. For the Kerr-Newman black hole case, when the angular frequency of the scalar perturbation satisfies $\omega<qQ/M$ and the product of the mass-to-charge ratios of the black hole and scalar perturbation satisfies $\frac{\mu}{q}\frac{M}{Q} > \frac{\sqrt{3 k^2+2} }{ \sqrt{k^2+2} },~k=\frac{a}{M}$, the extremal Kerr-Newman black hole is superradiantly stable under charged massive scalar perturbation.
\end{abstract}

\maketitle

\section{Introduction}
Superradiance is a well-known phenomenon in black hole physics\cite{Brito:2015oca}. When an ingoing scalar wave is scattered by a rotating or charged black hole, the outgoing wave may be amplified when the angular frequency of the scalar wave $\omega$ satisfies
\bea
0<\omega< m\Omega_H+q\Phi_H,
\eea
where $\Omega_H$ and $\Phi_H$ are respectively the angular velocity and electromagnetic potential of black hole horizon, $q$ and $m $ are respectively the charge and azimuthal quantum number of the ingoing scalar wave.

The analysis of (in)stability of the  black hole geometry is an important issue from both theoretical and astrophysical viewpoints. Due to superradiance, rotating or charged black holes suffer from superradiant instability. When there is a mirror outside the black hole horizon, the superradiant scalar wave may be scattered back and forth and the rotating or (and) electromagnetic energy of the black hole will be extracted continuously, which leads to the instability of the black hole. This mechanism was first discussed by Press and Teukolsky \cite{PTbomb} and is dubbed black hole bomb, which has been studied extensively in the literature \cite{Cardoso:2004nk,Dias:2018zjg,Hod:2016kpm,Hod:2016rqd,Rosa:2012uz,Hod:2013fvl,Hod:2014dda,Hod:2013nn,Hod:2009cp,Rosa:2009ei}. It is interesting that four-dimensional rotating Kerr  and Kerr-Newman (KN) black holes proved to be superradiantly unstable against massive scalar perturbation for some chosen parameters, where the mass of the scalar field provides a natural mirror\cite{Detweiler:1980uk,Zouros:1979iw,Strafuss:2004qc,Konoplya:2006br,Dolan:2007mj,Cardoso:2011xi,Dolan:2012yt,Hod:2012zza,Hod:2016iri,Degollado:2018ypf,Huang:2019xbu,Hod:2014pza,
Furuhashi:2004jk,Hod:2016bas,Huang:2016qnk,Huang:2018qdl}. For Kerr black hole case, in the regimes $M\mu\ll 1$ and $M\mu\gg 1$, where $M$ and $\mu$  are respectively the masses of the black hole and the perturbation field, the studies suggested that the instability growth rate is maximal for extremal black holes \cite{Detweiler:1980uk,Zouros:1979iw,Dolan:2007mj}.

The extremal black holes are special because geometrically their inner and outer horizons coincide and thermodynamically their temperatures are zero. Theoretically, it would be natural and interesting to study the extremal black holes from the viewpoint of the superradiant (in)stability, which is less studied in the literature. The four-dimensional extremal charged Reissner-Nordstrom black holes proved to be superradiantly stable under charged massive scalar perturbation\cite{Hod:2013eea}. Recently, the same stability properties have proved to hold for five or six-dimensional extremal Reissner-Nordstrom black holes and have been conjectured to hold for $D-$dimensional extremal Reissner-Nordstrom black holes\cite{Huang:2021dpa}. For extremal Kerr black holes, in the regime $M\mu\sim 1$, it is found that the superradiant instability growth rate decreases with increasing angular quantum number $l$ of the scalar perturbation field.  The dynamics of massive vector and tensor perturbation fields on a generic class of  extremal and near-extremal static black hole spacetimes has also been studied recently \cite{Cardoso:2019mes,Ueda:2018xvl}.

It is known that the superradiant (in)stability  property of  a black hole under perturbation depends on the value of parameters of the black hole and the perturbation filed. In this paper, we will analytically study the superradiantly stable parameter spaces for extremal Kerr and Kerr-Newman black holes under massive scalar perturbation. This theoretical study will be a complementary work for previous studies of the superradiant instability of the same systems. The superradiant stability of extremal rotating black holes is also important for high-energy particle physics in astrophysics.

The paper is organized as follows. In Section II, we give a description of the extremal Kerr black hole and scalar perturbation. By analysing the effective potential experienced by the scalar perturbation in the extremal Kerr black hole background, we find the superradiantly stable parameter space for the system consisting of the extremal Kerr black hole and the scalar perturbation. In Section III, a similar analysis is done for the extremal Kerr-Newman case. Section IV is devoted to the summary.

\section{Extremal Kerr black hole with a massive scalar perturbation}
In this section we describe the system we are interested in. The 4-dimensional Kerr black hole is stationary and axially symmetric. The corresponding metric in Boyer-Lindquist coordinates \cite{Boyer:1966qh} is
\begin{equation}
\begin{aligned}
d s^{2}=& \frac{\Delta}{\rho^{2}}\left(d t-a \sin ^{2} \theta d \phi\right)^{2}+\frac{\rho^{2}}{\Delta} d r^{2}+\rho^{2} d \theta^{2} \\
&+\frac{\sin ^{2} \theta}{\rho^{2}}\left[a d t-\left(r^{2}+a^{2}\right) d \phi\right]^{2},
\end{aligned}
\end{equation}
where
\begin{equation}
\Delta=r^{2}-2 M r+a^{2}, \rho^{2}=r^{2}+a^{2} \cos ^{2} \theta.
\end{equation}
$M$ is the mass of the black hole and $a$ is the angular momentum per unit mass of the black hole. The inner and outer horizons of the Kerr black hole are

\begin{equation}r_{\pm}=M \pm \sqrt{M^{2}-a^{2}}.\end{equation}

For extremal Kerr black holes, $M=a$, the inner and outer horizons coincide,
\begin{equation}r_{+}=r_{-}=r_h=M.\end{equation}
The angular velocity of the extremal Kerr black hole horizon is
\begin{equation}\Omega_{H}=\frac{a}{r_{h}^{2}+a^{2}}=\frac{1}{2 a}.\end{equation}

The dynamics of a massive scalar field perturbation $\Psi$ is governed by the covariant Klein-Gordon equation, which is
\begin{equation}\left(\nabla^{v} \nabla_{v}-\mu^{2}\right) \Psi=0.\end{equation}
The solution of the above differential equation with definite angular frequency $\omega$ can be decomposed as
\begin{equation}\Psi(t, r, \theta, \phi)=\sum_{l, m} R_{l m}(r) S_{l m}(\theta) e^{i m \phi} e^{-i \omega t},\end{equation}
where $l$ and $m$ respectively represent the spheroidal harmonic index and the azimuthal harmonic index of the mode. $R_{l m}$ is the radial function and $S_{l m}(\theta)$ is the angular function. The angular function is  the standard spheroidal harmonics and it satisfies the following angular equation
\begin{equation}\begin{aligned}
\frac{1}{\sin \theta} & \frac{d}{d \theta}\left(\sin \theta \frac{d S_{l m}}{d \theta}\right) \\
&+\left[J_{l m}+\left(\mu^{2}-\omega^{2}\right) a^{2} \sin ^{2} \theta- \frac{m^{2}}{\sin ^{2} \theta}\right] S_{l m}=0,
\end{aligned}\end{equation}
where $J_{l m}$ is the eigenvalue and the asymptotic expansion of it in various limits can be found in\cite{Berti:2005gp}.
There is a lower bound for $J_{l m}$ \cite{Hod:2015cqa},
\begin{equation}\label{Jlm}
J_{l m}>m^2-a^{2}\left(\mu^{2}-\omega^{2}\right),
\end{equation}
which is an important inequality in our following discussion.

For radial function $R_{lm}$, it satisfies the following radial equation
\begin{equation}\label{rad-eq-k}
\Delta \frac{d}{d r}\left(\Delta \frac{d R_{l m}}{d r}\right)+U_K R_{l m}=0,
\end{equation}
where
\begin{equation}
\begin{split}
&U_{K}=\left[\omega\left(r^{2}+a^{2}\right)-m a\right]^{2}\\
&+\Delta\left[2 m a \omega-\mu^{2}\left(r^{2}+a^{2}\right)-J_{l m}\right].
\end{split}
\end{equation}

Define the tortoise coordinate $r_*$ by
\begin{equation}\frac{d r_{*}}{d r}=\frac{r^{2}+a^{2}}{\Delta},\end{equation}
and a new radial function $\tilde{R}_{lm}=\sqrt{r^{2}+a^{2}} R_{l m}$, the radial equation \eqref{rad-eq-k} can be rewritten as
\begin{equation}
\frac{d^{2} \tilde{R}_{l m}}{d r_{*}^{2}}+\tilde{U} \tilde{R}_{l m}=0.
\end{equation}
In order to discuss the superradiance modes, the physical boundary conditions for the radial function  are ingoing wave at the black hole horizon and exponentially decaying bound state at spatial infinity. The asymptotic solutions of the above equation at the boundaries are
\begin{equation}\begin{array}{l}
r \rightarrow+\infty\left(r_{*} \rightarrow+\infty\right), \quad \tilde{R}_{l m} \sim e^{-\sqrt{\mu^{2}-\omega^{2}} r_{*}}, \\
r \rightarrow r_{h}\left(r_{*} \rightarrow-\infty\right), \quad \tilde{R}_{l m} \sim e^{-i\left(\omega-m \Omega_{H}\right) r_{*}}.
\end{array}\end{equation}
Apparently, the exponentially decaying bound state condition requires
the following inequality
\begin{equation}\label{bound-state-con-k}
\omega^{2}<\mu^{2}.
\end{equation}
For Kerr black holes, the superradiant condition is
\begin{equation}
\omega<\omega_c=m \Omega_{H}.
\end{equation}

To analyse the superradiant stability of extremal Kerr black hole, we define a new radial function
$\psi= \Delta^{1 / 2} R_{lm}$ and the radial equation\eqref{rad-eq-k} can be rewritten as a Schrodinger-like equation
\begin{equation}
\frac{d^{2} \psi}{d r^{2}}+\left(\omega^{2}-V_{K}\right) \psi=0,
\end{equation}
where the effective potential $V_K$ is
\bea
V_{K}=\omega^{2}-\frac{U_{K}}{\Delta^{2}}.
\eea
For the superradiant modes, we will find the parameter space of the system where there is no trapping well outside the black hole horizon for the effective potential. In this parameter space, the system is superraiantly stable.

Let's analyse the asymptotic behaviors of the effective potential $V_K$,
\begin{equation}
V_{K}(r \rightarrow +\infty) \rightarrow \mu ^2+\frac{2 a (\mu ^2-2 \omega ^2)}{r}+O(\frac{1}{r ^2}),\end{equation}
\begin{equation}
V_{K}\left(r \rightarrow r_{h}\right) \rightarrow-\infty,
\end{equation}
\begin{equation}
V_{K}^{\prime}(r \rightarrow +\infty) \rightarrow \frac{2 a( 2\omega ^2- \mu ^2)}{r^2}+O(\frac{1}{r ^3}).
\end{equation}
In order that there is no trapping potential well for $V_K$ outside the horizon, we need the condition, $V_{K}^{\prime}(r \rightarrow \infty)<0$, i.e.
\begin{equation}\label{o-mu}
\omega^{2}<\frac{\mu^{2}}{2}.
\end{equation}

\subsection{The superradiant stability analysis of extremal Kerr black holes}
In this part, by analysing the shape of the effective potential $V_K$, we will determine the parameter space where the system of extremal Kerr black hole and massive scalar perturbation is superradiantly stable. As mentioned before, we need the parameter space where there is no trapping potential well outside the horizon.

We define a new coordinate $z$, $z=r-{r_h}$. Then the explicit expression of the derivative of the effective potential $V$ is
\begin{equation}
\begin{split}
V'( r) =V'( z) &=\frac{Ar^4+Br^3+Cr^2+Dr}{-\varDelta ^3}\\
&=\frac{A_1z^4+B_1z^3+C_1z^2+D_1z}{-\varDelta ^3}.
\end{split}
\end{equation}
The relations between the above coefficients are
\begin{equation}
\begin{split}
{A_1} &= A,\\
{B_1} &= B+4r_hA_1,\\
{C_1} &= C+3r_hB_1-6r_{h}^{2}A_1,\\
{D_1} &= D+4r_{h}^{2}A_1-3r_{h}^{2}B_1+2r_hC_1.
\end{split}
\end{equation}
 In the following, we will use $f(z)$ to denote the numerator of the derivative of the effective potential $V$, which is a polynomial in z,
 \bea\label{f-kerr}
 f(z)=A_1z^3+B_1z^2+C_1z+D_1.
  \eea
 By analyzing the signs of roots of the equation $f(z) = 0$, we can determine whether there are potential wells outside the horizon.
 The equation $f(z) = 0$ has three roots: $z_1$, $z_2$ and $z_3$. According to Vieta theorem, we have the following relation:
\begin{eqnarray}
\label{z123}
z_1z_2z_3= -\frac{D_1}{A_1},\\ \label{z12}
z_1z_2+z_1z_3+z_2z_3=\frac{C_1}{A_1},\\ \label{z1}
z_1+z_2+z_3=-\frac{B_1}{A_1}.
\end{eqnarray}
According to the asymptotic behaviors of the effective potential, we know that there exists at least one positive root, denoted by $z_1$, outside the horizon, i.e.
\begin{equation}\label{z1-p}
z_1>0.
\end{equation}

In the following section, we will analyze the  distribution of the roots $z_2$, $z_3$ of the equation $f(z)=0$ and find the interesting parameter region where it is superradiantly stable for the system of an extremal Kerr black hole and massive scalar perturbation.

\subsection{Coefficient analysis of extremal Kerr black holes}

In this subsection, we will study the roots of equation $f(z)=0$ by analyzing the four coefficients in \eqref{f-kerr}.  The four coefficients can be written explicitly as following,
\begin{equation}
\begin{split}
{A_1}\equiv{A_K} &= 2 a (\mu ^2-2 \omega ^2),\\
{B_1}\equiv{B_K} &= 2 (2 a^2 (\mu ^2-4 \omega ^2)+J_{{lm}}),\\
{C_1}\equiv{C_K} &= 12 a^2 \omega  (m-2 a \omega ),\\
{D_1}\equiv{D_K} &= -4 a^2 (m-2 a \omega )^2.
\end{split}
\end{equation}
It is easy to see that ${{{D}_{K}}}< 0$. According to the condition \eqref{o-mu}, we have ${{{A}_{K}}}> 0$.
From equation \eqref{z123}, we can deduce that $z_2$ and $z_3$ are both either positive or negative.

In the following, we will find the parameter region where $B_k$ is positive. Then the roots $z_2$ and $z_3$ are both negative and there is only one maximum for the effective potential outside the black hole horizon. The Kerr black hole and scalar perturbation system is superradiantly stable.

\begin{figure}[htbp]
\centering
\subfigure[]{
\begin{minipage}{6cm}\centering
\includegraphics[scale=0.25]{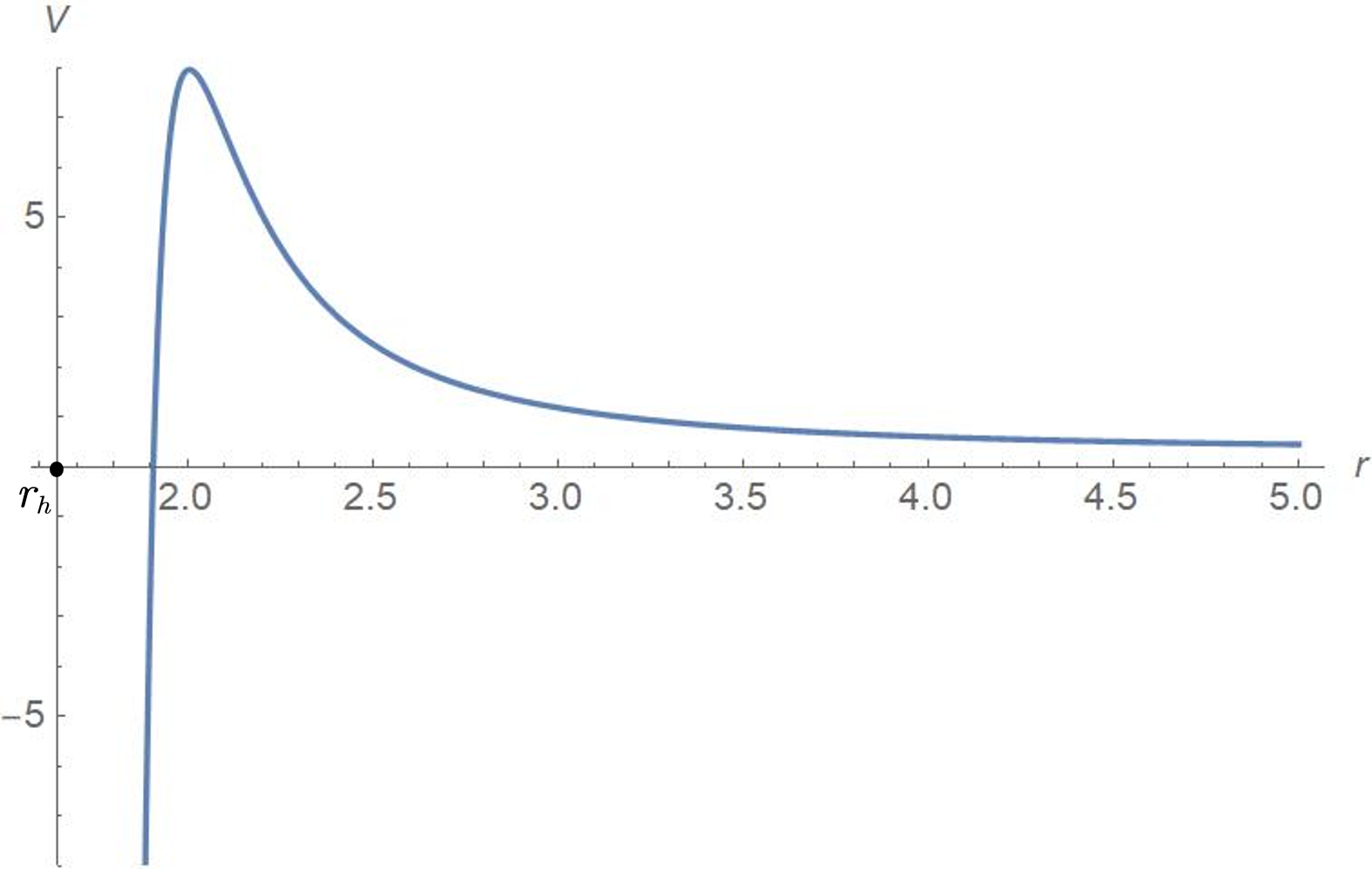}
\end{minipage}}
\subfigure[]{
\begin{minipage}{6cm}\centering
\includegraphics[scale=0.25]{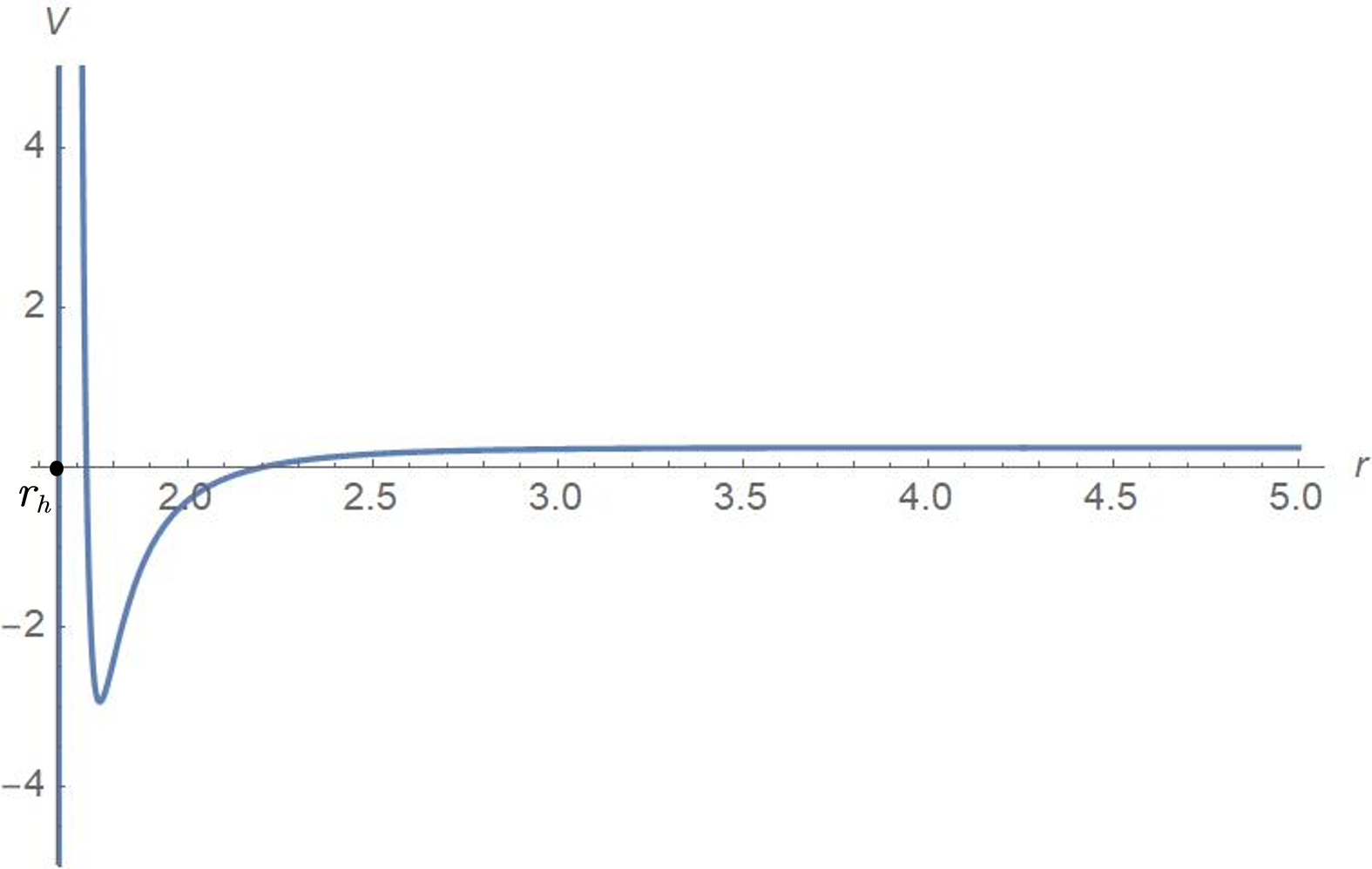}
\end{minipage}}
\caption{Two typical types of the effective potential $V$. (a) is a superradiantly stable. The parameters of the scalar field are chosen as $m=1,\omega=0.214, \mu=0.478$. The mass parameter of the extremal Kerr black hole is $M=a=1.650$. (b) is superradiantly unstable. The parameters are taken as $m=1, \omega=0.302,\mu=0.465,M=a=1.650$. }
\label{Kerr}
\end{figure}

According to the lower bound for the eigenvalue of the angular equation $J_{lm}$ in \eqref{Jlm}, we have
\begin{equation}
{B_K}=2 (2 a^2 (\mu ^2-4 \omega ^2)+J_{{lm}})
>2a^2 (\mu ^2-7 \omega ^2)+2m^2.
\end{equation}
The superradiance condition for our extremal Kerr black hole case is
\bea
m \Omega_H=\frac{m}{2 a}>\omega.
\eea
So given the superradiance condition the coefficient $B_K$ satisfies
\bea
B_K>2a^2\mu^2-6a^2\omega^2.
\eea
If the mass $\mu$  and angular frequency $\omega$ of the massive scalar perturbation satisfy the following inequality
\begin{equation}
\mu >\sqrt{3} \omega,
\end{equation}
$B_K$ is positive. Under the above condition, there is no trapping potential well for the effective potential, the system of extremal Kerr black hole and scalar perturbation is superradiantly stable. With two sets of chosen parameters of the extremal Kerr black hole and scalar field,
we illustrate two typical types of the effective potentials in the extremal Kerr black hole case in Fig.\eqref{Kerr}

\section{Extremal Kerr-Newman black hole with a charged massive scalar perturbation}
In this section, we consider a system consisting of a charged massive scalar field minimally coupled to an extremal Kerr-Newman black hole.
For simplicity, some of the symbols in this section are the same as that of Kerr black hole system.

The background spacetime is described by the Kerr-Newman line element, which, in the standard Boyer-Lindquist coordinates, is given by
\begin{equation}\begin{aligned}
d s^{2}=&-\frac{\Delta}{\rho^{2}}\left(d t-a \sin ^{2} \theta d \phi\right)^{2}+\frac{\rho^{2}}{\Delta} d r^{2} \\
&+\rho^{2} d \theta^{2}+\frac{\sin ^{2} \theta}{\rho^{2}}\left[\left(r^{2}+a^{2}\right) d \phi-a d t\right]^{2},
\end{aligned}\end{equation}
where
\begin{equation}
\rho^{2} \equiv r^{2}+a^{2} \cos ^{2} \theta, \Delta \equiv r^{2}-2 M r+a^{2}+Q^{2},
\end{equation}
where $Q$, $M$ and $a$ respectively represent the charge, mass and angular momentum per unit mass of the Kerr-Newman black hole.

For extremal Kerr-Newman black hole, the inner and outer horizons coincide,
\begin{equation}
r_{+}=r_{-}=r_h=M,
\end{equation}
and we have
\begin{equation}
M=\sqrt{a^{2}+Q^{2}}.
\end{equation}
The background electromagnetic potential in Kerr-Newman black hole is
\begin{equation}
A_{\mu}=\left(-\frac{Q r}{\rho^{2}}, 0,0, \frac{a Q r \sin ^{2} \theta}{\rho^{2}}\right).
\end{equation}

The dynamics of a charged massive scalar field $\Psi$ in the Kerr-Newman spacetime is described by the following Klein-Gordon wave equation
\begin{equation}
\left(\nabla^{v}-i q A^{v}\right)\left(\nabla_{v}-i q A_{v}\right) \Psi=\mu^{2} \Psi,
\end{equation}
where $\nabla^{v}$ stands for the covariant derivative in the Kerr-Newman geometry,  $\mu$ and $q$ represent the mass and charge of the scalar field.

The solution of the above equation with definite angular frequency can be decomposed as
\begin{equation}\Psi(t, r, \theta, \Phi)=\sum_{l m} R_{l m}(r) S_{l m}(\theta) e^{i m \phi} e^{-i \omega t}.\end{equation}
The angular and radial parts of the Klein-Gordon equations share the same forms with the extremal Kerr black hole case. The only difference is that in the Kerr-Newman case,  $U_K$ in radial equation \eqref{rad-eq-k} is replaced by
\begin{equation}\begin{aligned}
U_{KN}=&\left[\omega\left(a^{2}+r^{2}\right)-a m-q Q r\right]^{2} \\
&+\Delta\left[2 a m \omega-J_{l m}-\mu^{2}\left(r^{2}+a^{2}\right)\right].
\end{aligned}\end{equation}
So in this case the radial equation of motion is
\begin{equation}\label{rad-eq-kn}
\Delta \frac{d}{d r}\left(\Delta \frac{d R_{l m}}{d r}\right)+U_{KN} R_{l m}=0.
\end{equation}
The angular equation in extremal Kerr-Newman case is the same as that in the extremal Kerr case. So the eigenvalues  $J_{l m}$ satisfy the same lower bound \eqref{Jlm}.

Define the tortoise coordinate $r_*$ as following
\begin{equation}
\frac{d r_{*}}{d r}=\frac{r^{2}+a^{2}}{\Delta},
\end{equation}
Here we consider classical scalar wave. The appropriate boundary conditions we are interested in at the horizon and spatial infinity are purely ingoing wave near the horizon and exponentially decaying wave at spatial infinity. With these boundary conditions, the radial function has the following asymptotic solutions
\begin{equation}\label{asym-kn}
R_{l m}(r) \sim\left\{\begin{array}{l}
e^{-i\left(\omega-\omega_{c}\right) r_{*}}, r_*\rightarrow -\infty (r \rightarrow r_h),\\
\frac{e^{-\sqrt{\mu^{2}-\omega^{2}} r}}{r}, r_*\rightarrow +\infty (r \rightarrow +\infty),
\end{array}\right.\end{equation}
where the critical frequency $\omega_c$ are defined as
\begin{equation}
\omega_{c}=m \Omega_{H}+q \Phi_{H},
\end{equation}
where $\Omega_H=\frac{a}{M^{2}+a^2}$ is the angular velocity of the horizon and $\Phi_{H}=\frac{Q M}{M^{2}+a^{2}}$ is the electric potential of the horizon.
From eq.\eqref{asym-kn}, we can also obtain that the superradiance condition is
\bea\label{super-cond-kn}
0<\omega<\omega_{c}=m \Omega_{H}+q \Phi_{H},
\eea
and the bound state (exponentially decaying) condition is
\begin{equation}\label{bound}
\omega^{2}<\mu^{2}.
\end{equation}

In order to analyse the superradiant stability of extremal Kerr-Newman black hole, we define a new radial function
\bea
\psi \equiv \Delta^{1/2} R_{lm},
\eea
and the radial equation of motion can be rewritten as a Schrodinger-like equation
\begin{equation}
\frac{d^{2} \psi}{d r^{2}}+\left(\omega^{2}-V_{KN}\right) \psi=0,
\end{equation}
where the effective potential is
\bea\label{eff-pot-kn}
 \quad V_{KN}=\omega^{2}-\frac{U_{KN}}{\Delta^{2}}.
\eea

The asymptotic behaviors of the effective potential at the event horizon and spatial infinity are
\begin{equation}
\begin{aligned}
V_{KN}\rightarrow &\frac{2 \mu ^2 \sqrt{a^2+Q^2}-4 \omega ^2 \sqrt{a^2+Q^2}+2 q Q \omega }{r}\\
&+\mu^2+O(\frac{1}{r ^2}), ~r \rightarrow +\infty,\\
V_{KN}\rightarrow&-\infty, ~r \rightarrow r_h.
\end{aligned}\end{equation}
The asymptotic behavior of the derivative of the effective potential at spatial infinity is
\begin{equation}
\begin{aligned}
V_{KN}^{\prime}&\rightarrow \frac{-2 \mu ^2 \sqrt{a^2+Q^2}+4 \omega ^2 \sqrt{a^2+Q^2}-2 q Q \omega }{r^2}\\
&+O(\frac{1}{r ^3}),~ r \rightarrow +\infty.
\end{aligned}\end{equation}
In order that there is no trapping potential well near spatial infinity, we need $V_{KN}^{\prime}(r \rightarrow +\infty)<0$, i.e.
\bea
-2 \mu ^2 \sqrt{a^2+Q^2}+4 \omega ^2 \sqrt{a^2+Q^2}-2 q Q \omega<0.
\eea
Given the condition \eqref{bound}, the above inequality holds if
\bea
2 \omega ^2 \sqrt{a^2+Q^2}-2 q Q \omega<0,
\eea
i.e.
\begin{equation}\label{KN-result-1}
\omega<\frac{q Q}{\sqrt{a^2+Q^2}}=\frac{qQ}{M}.
\end{equation}
When the above inequality holds, we have
\begin{equation}
V_{KN}^{\prime}(r \rightarrow +\infty)<0.
\end{equation}
Then, according to the asymptotic behaviors of $V_{KN}$ at the event horizon and spatial infinity, we know that there is at least one maximum for $V_{KN}$
when $r>r_h$.
\subsection{Coefficient analysis of the Extremal Kerr-Newman black holes}
In order to find the superradiant stable parameter space, we need to prove further that there is no trapping potential well outside the horizon of the extremal Kerr-Newman black hole for the effective potential \eqref{eff-pot-kn}.
Following the method discussed in Section II.1, we replace the variable $r$ with $z=r-r_h$ and the numerator of the derivative of the effective potential is a polynomial of $z$,\eqref{f-kerr}. The three roots are also denoted by $z_1,z_2,z_3$. The four coefficients $A_1,B_1,C_1,D_1$ for the extremal Kerr-Newman black hole case are as follows,
\begin{equation}
\begin{aligned}\label{2}
&{A_{1}}\equiv{A_{KN}} = 2 [\omega  (q Q-2 \omega  \sqrt{a^2+Q^2})+\mu ^2 \sqrt{a^2+Q^2}],\\
&{B_{1}}\equiv{B_{KN}} = -2 (-2 a^2 \mu ^2-J_{lm}+q^2 Q^2-\mu ^2 Q^2) \\
&+12 q Q \omega  \sqrt{a^2+Q^2}-2 \omega ^2 (8 a^2+6 Q^2),\\
&{C_{1}}\equiv{C_{KN}} = 6 (-q^2 Q^2 \sqrt{a^2+Q^2}-a m q Q) \\
&+6 \omega  (2 a m \sqrt{a^2+Q^2}+4 a^2 q Q+3 q Q^3) \\
&+6 \omega ^2 (-4 a^2 \sqrt{a^2+Q^2}-2 Q^2 \sqrt{a^2+Q^2}),\\
&{D_{1}}\equiv{D_{KN}} = -4 \omega ^2 (4 a^4+4 a^2 Q^2+Q^4)+4 \omega(4 a^3 m \\
&+4 a^2 q Q \sqrt{a^2+Q^2}+2 q Q^3 \sqrt{a^2+Q^2}+2 a m Q^2)\\
&-4 (a^2 m^2+2 a m q Q \sqrt{a^2+Q^2}+a^2 q^2 Q^2+q^2 Q^4).
\end{aligned}
\end{equation}
According to the asymptotic analysis of $V_{KN}$, we know that there is at least one root for $V'_{KN}=0$ when $z>0$. This root can be denoted by $z_1$.

According to the condition \eqref{KN-result-1}, we have ${{{A}_{KN}}}>0$.
The coefficient ${D_{KN}}$ can be rewritten as
\bea
D_{KN}=-(\omega-\omega_c)^2,
\eea
where
\begin{equation}
\begin{aligned}\nn
\omega _c=\frac{q Q \sqrt{a^2+Q^2}+a m}{2 a^2+Q^2},
\end{aligned}
\end{equation}
is the critical angular frequency in the superradiance condition \eqref{super-cond-kn}. So given the superradiance condition, we have
\bea
D_{KN}<0.
\eea
From equation \eqref{z123}, we obtain that $z_2,z_3$ are both negative or positive. If $z_2,z_3$ are both negative, there is only one extreme (maximum) for the effective potential outsider the horizon which is located at $z_1$. Then $V_{KN}$ has no potential well outside the horizon and the Kerr-Newman black hole are superradiantly stable.
According to Vieta theorem \eqref{z1}, the positivity condition for $B_{KN}$ is a sufficient condition for negativity of $z_2,z_3$. In the next, we will find a positivity condition for $B_{KN}$.

The coefficient $B_{KN}$ is
\bea
&&{B_{KN}} = \mu ^2 \left(2 a^2+Q^2\right)+2J _{{lm}}\nn\\
&&-2 \left(-6 q Q \omega  \sqrt{a^2+Q^2}+8 a^2 \omega ^2+q^2 Q^2+6 Q^2 \omega ^2\right).\nn
\eea
Given the lower bound on $J_{lm}$ in \eqref{Jlm},
we can obtain
\begin{equation}
\begin{aligned}
&{B_{KN}}>12 q Q \omega  \sqrt{a^2+Q^2}+\mu ^2 \left(3 a^2+2 Q^2\right)\\
&-15 a^2 \omega ^2+2 m^2-2 q^2 Q^2-12 Q^2 \omega ^2.
\end{aligned}
\end{equation}
The positivity of the right side of the above inequality is equivalent to
\begin{equation}
\begin{aligned}\label{KN-mu2}
&\mu^2 >\\
&\frac{15 a^2 \omega ^2-12 q Q \omega  \sqrt{a^2+Q^2}-2 m^2+2 q^2 Q^2+12 Q^2 \omega ^2}{3 a^2+2 Q^2}.
\end{aligned}
\end{equation}
The numerator of the right side of the above inequality can be treated as a quadratic function  of  $\omega$, which is defined as follows
\begin{equation}
\begin{aligned}
&g(\omega)= 15 a^2 \omega ^2-12 q Q \omega  \sqrt{a^2+Q^2}\\
& -2 m^2+2 q^2 Q^2+12 Q^2 \omega ^2.
\end{aligned}
\end{equation}
It is easy to see the symmetric axis of $g(\omega)$ is at
\bea
\omega_{A}=\frac{2 q Q \sqrt{a^2+Q^2}}{5 a^2+4 Q^2}.
\eea
According to the previous asymptotic analysis the effective potential \eqref{KN-result-1}, we obtain a constraint on $\omega$, i.e. $0<\omega<\frac{q Q}{\sqrt{a^2+Q^2}}$.
Because the following inequality
\begin{equation}
\begin{aligned}
\frac{q Q}{\sqrt{a^2+Q^2}}>\frac{q Q}{\frac{a^2}{4 \sqrt{a^2+Q^2}}+\sqrt{a^2+Q^2}}\\
=2 \frac{2 q Q \sqrt{a^2+Q^2}}{5 a^2+4 Q^2}=2\omega_A,
\end{aligned}
\end{equation}
we know when $0<\omega<\frac{q Q}{\sqrt{a^2+Q^2}}$ the maximum of $g(\omega)$ is $g(\frac{q Q}{\sqrt{a^2+Q^2}})$. A sufficient condition for the inequality \eqref{KN-mu2} is
\begin{equation}
\begin{aligned}
\mu^2>\frac{g(\frac{q Q}{\sqrt{a^2+Q^2}})+2m^2}{3 a^2+2 Q^2}=\frac{\frac{q^2 Q^2 \left(5 a^2+2 Q^2\right)}{a^2+Q^2}}{3 a^2+2 Q^2}.
\end{aligned}
\end{equation}
For the extremal Kerr-Newman black hole, we have $\sqrt{a^2+Q^2}=M$.  The above inequality is equivalent to
\begin{equation}
\label{KN-result-2}
\frac{\mu}{q}\frac{M}{Q} > \frac{\sqrt{3 k^2+2} }{ \sqrt{k^2+2} },~~k=\frac{a}{M}.
\end{equation}
When the above inequality holds, $B_{KN}>0$ and $z_2,z_3$ are both negative. There is no trapping potential well outside the black hole horizon. The extremal Kerr-Newman black hole and scalar perturbation system is superradiantly stable.

\begin{figure}[htbp]
\centering
\subfigure[]{
\begin{minipage}{6cm}\centering
\includegraphics[scale=0.25]{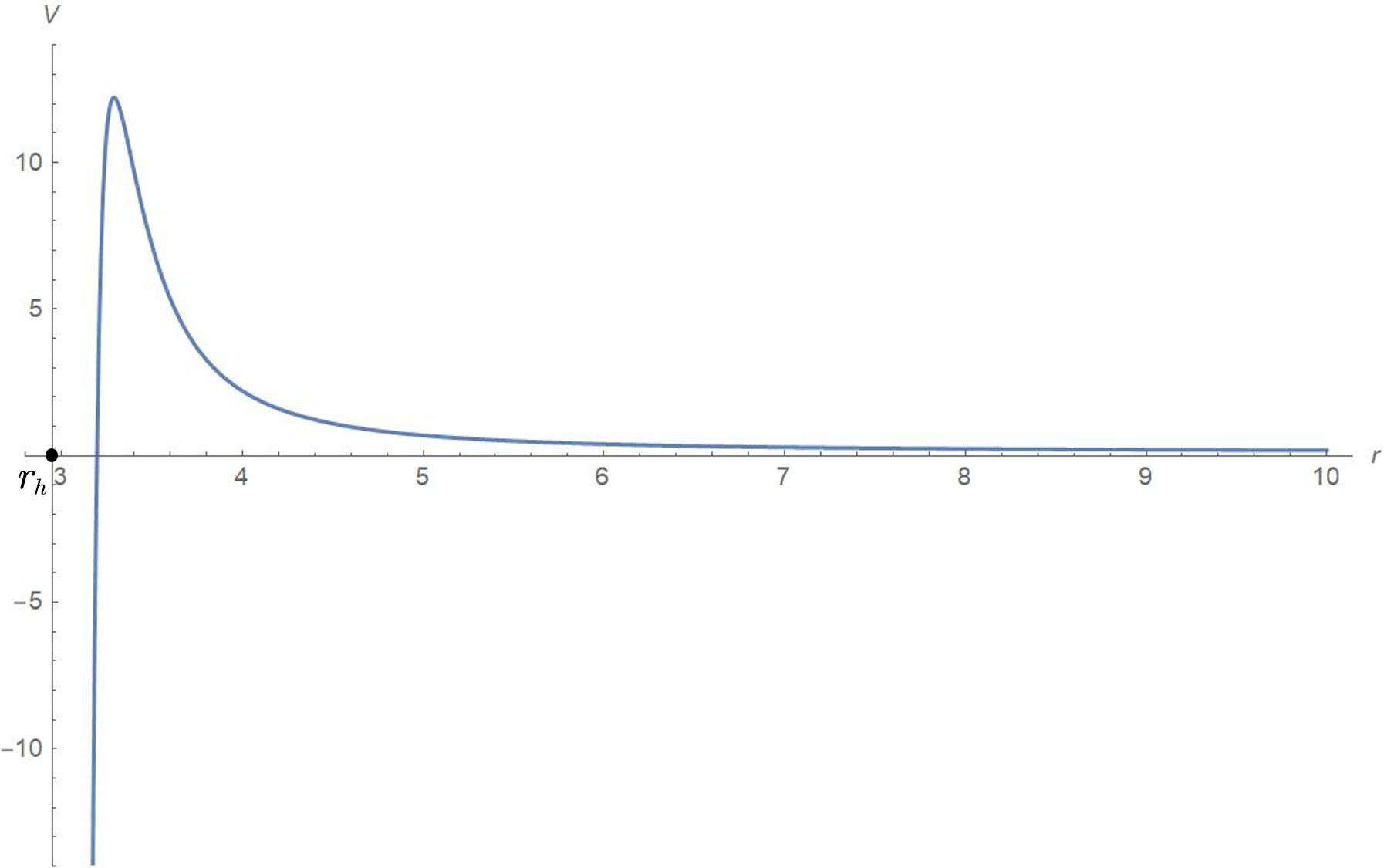}
\end{minipage}}
\subfigure[]{
\begin{minipage}{6cm}\centering
\includegraphics[scale=0.25]{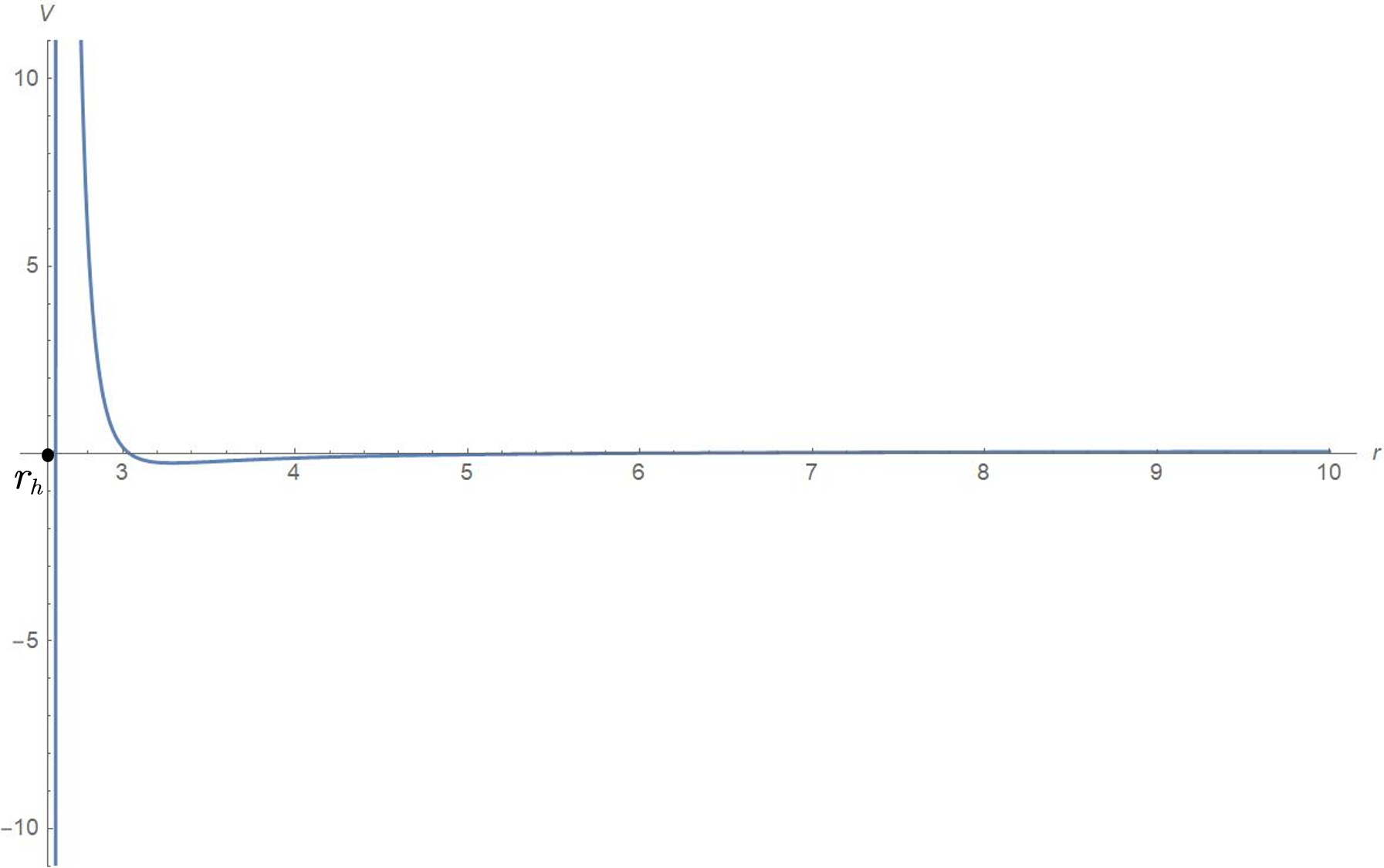}
\end{minipage}}
\caption{Two typical types of the effective potential $V$. (a) is a superradiantly stable. The parameters of the scalar field are chosen as $m=1,\omega=0.266, \mu=0.336, q=0.290$. The parameters of the extremal KN black hole are $a=0.400, Q=2.920$. (b) is superradiantly unstable. The parameters are taken as $m=1, \omega=0.285,\mu=0.286,q=0.271,a=2.010, Q=1.590$. }
\label{kn}
\end{figure}

Finally, two typical types of effective potential in this case are shown in Fig.\eqref{kn}. For the superradiantly stable case, one can check that $\frac{\mu M}{q Q}=1.169>\sqrt{\frac{3k^2+2}{k^2+2}}=1.009$ and
$\omega=0.266<\frac{qQ}{M}=0.287$. For the superradiantly unstable case, one can check $\omega=0.285>\frac{qQ}{M}=0.168$.

\section{Summary}
In this paper we analytically study the superradiant stable regimes for four-dimensional extremal Kerr and Kerr-Newman black holes under scalar perturbation. The logic in our discussion is that given the existence of superradiant modes and a bound state condition at spatial infinity, there is no trapping potential well for the effective potential experienced by the scalar perturbation.

The equation of motion of the scalar perturbation is governed by the covariant Klein-Gordon equation in the black hole background. The equation of motion can be separated into angular and radial parts. From the radial part of the equation of motion, we can obtain the effective potential experienced by the scalar field. By requiring there is no potential well outside the horizon of the black hole, we determine the superradiant stable parameter space for the black hole and scalar perturbation system. In the Kerr black hole case, it is found that all the extremal Kerr black holes are superradiantly stable when the angular frequency and proper mass of the scalar perturbation satisfy the inequality $\omega<\mu/\sqrt{3}$. The extremal Kerr-Newman black holes are superradiantly stable under charged massive scalar perturbation when the angular frequency of the scalar perturbation satisfies $\omega<qQ/M$ and the product of the mass-to-charge ratios of the black hole and perturbation satisfy $\frac{\mu}{q}\frac{M}{Q} > \frac{\sqrt{3 k^2+2} }{\sqrt{k^2+2}},~k=\frac{a}{M}$. Our results provide further knowledge on the superradiant stability property of Kerr and Kerr-Newman black holes under scalar perturbation.

\begin{acknowledgements}
This work is partially supported by Guangdong Major Project of Basic and Applied Basic Research (No. 2020B0301030008), Science and Technology Program of Guangzhou (No. 2019050001) and Natural Science Foundation of Guangdong Province (No. 2020A1515010388).
\end{acknowledgements}

\end{document}